\documentclass[reprint,twocolumn,prl,superscriptaddress,showkeys,floatfix]{revtex4-1}
\usepackage{graphicx}
\usepackage{dcolumn}
\usepackage{bm}
\usepackage{color}
\usepackage{array}
\usepackage{multirow}
\usepackage{fix-cm}
\usepackage{pbox}
\usepackage{amssymb}
\usepackage{amsmath}
\usepackage{wasysym}
\usepackage{float}
\usepackage{sidecap}
\usepackage{color}
\usepackage{sidecap}
\usepackage[normalem]{ulem}
\usepackage[resetlabels]{multibib}
\newcites{supp}{Supplementary Material}
\usepackage{ifthen}
\newboolean{printfigures}
\setboolean{printfigures}{true}

\renewcommand{\sout}[1]{}

\definecolor{orange}{rgb}{1,0.5,0}

\renewcommand{\thesection}{\arabic{section}}
\renewcommand{\thetable}{\arabic{table}}
\renewcommand{\thefigure}{\arabic{figure}}
\makeatletter
\renewcommand\p@subsubsection{}
\makeatother
\renewcommand{\thesubsubsection}{\arabic{section}.\arabic{subsubsection}}
\renewcommand{\thesubsection}{\arabic{section}.\arabic{subsection}}

\begin{document}

\title{The Structure of Cholesterol in Lipid Rafts}

\author{Laura~Toppozini}
\affiliation{Department of Physics and Astronomy, McMaster
University, Hamilton, Ontario, Canada}

\author{Sebastian~Meinhardt}
\affiliation{KOMET 331, Institute of Physics, Johannes
Gutenberg-Universit\"at Mainz, Mainz, Germany}

\author{Clare~L.~Armstrong}
\affiliation{Department of Physics and Astronomy, McMaster
University, Hamilton, Ontario, Canada}

\author{Zahra~Yamani}
\affiliation{Canadian Neutron Beam Centre, Chalk River, Ontario,
Canada}

\author{Norbert~Ku\v{c}erka}
\affiliation{Canadian Neutron Beam Centre, Chalk River, Ontario,
Canada}

\author{Friederike~Schmid}\email{friederike.schmid@uni-mainz.de}
\affiliation{KOMET 331, Institute of Physics,
Johannes Gutenberg-Universit\"at Mainz, Mainz, Germany}

\author{Maikel~C.~Rheinst\"adter}\thanks{rheinstadter@mcmaster.ca}
\affiliation{Department of Physics and Astronomy, McMaster
University, Hamilton, Ontario, Canada} \affiliation{Canadian Neutron
Beam Centre, Chalk River, Ontario, Canada}

\date{\today}

\begin{abstract}
Rafts, or functional domains, are transient nano- or mesoscopic
structures in the plasma membrane and are thought to be essential
for many cellular processes such as signal transduction, adhesion,
trafficking and lipid/protein sorting. Observations of these
membrane heterogeneities have proven challenging, as they are
thought to be both small and short-lived. With a combination of
coarse-grained molecular dynamics simulations and neutron
diffraction using deuterium labeled cholesterol molecules we observe
raft-like structures and determine the ordering of the cholesterol
molecules in binary cholesterol-containing lipid membranes. From coarse-grained
computer simulations, heterogenous membranes structures were
observed and characterized as small, ordered domains. Neutron
diffraction was used to study the lateral structure of the
cholesterol molecules. We find pairs of strongly bound cholesterol
molecules in the liquid-disordered phase, in accordance with the
umbrella model. Bragg peaks corresponding to ordering of the
cholesterol molecules in the raft-like structures were observed and
indexed by two different structures: a monoclinic structure of
ordered cholesterol pairs of alternating direction in equilibrium
with cholesterol plaques, i.e., triclinic cholesterol bilayers.
\end{abstract}


\maketitle
The liquid-ordered ($l_o$) phase of membranes in the presence of
cholesterol was brought to the attention of the life science
community in 1997 when Simons and Ikonen \cite{Simons:1997} proposed
the existence of so-called rafts in biological membranes. Rafts were
thought to be small, molecularly organized units, providing local
structure in fluid biological membranes and hence furnishing
platforms for specific biological functions
\cite{Simons:1997,Simons:2000,Engelmann:2005,Niemela:2007,Pike:2009,Lingwood:2010,Eggeling:2009,Goot:2001,Lenne:2009,Simons:2010}.
These rafts were supposed to be enriched in cholesterol making them
more ordered, thicker and, thus, appropriate anchoring places for
certain acylated and hydrophobically-matched integral membrane
proteins. The high levels of cholesterol in these rafts led to the
proposal that rafts are local manifestations of the $l_o$ phase,
although in most cases the nature of the lipid ordering and the
phase state were not established in cells, nor in most model
membrane studies~\cite{Mouritsen:2010,Simons:2010,Rheinstadter:2013}.

Rafts are generally interpreted as some kind of super-particles
floating around in an otherwise structureless liquid membrane.
However, early work in the physical chemistry of lipid bilayers
pointed to the possibility of dynamic heterogeneity
\cite{Dibble:1996,Mouritsen:1993,Mouritsen:1994,Mouritsen:1997} in
thermodynamic one-phase regions of binary systems. The sources of
dynamic heterogeneity are cooperative molecular interactions and
thermal fluctuations that lead to density and compositional
fluctuations in space and time.

A number of ternary phase diagrams have been determined for systems
involving cholesterol and two different lipid species. Usually these
systems contain a lipid with a high melting point, such as a
long-chain saturated phospholipid, and a lipid with a low melting
point, such as sphingolipids or unsaturated phospholipids
\cite{Marsh:2010}, resulting in the observations of
micrometer-sized, thermodynamically stable domains
\cite{Niemela:2007,Risselada:2008,Berkowitz:2009,Bennett:2013,Heberle:2013}.
Much less work has been done on cholesterol/lipid binary mixtures,
which although seemingly simpler, have proven to be more difficult to
study. Evidence for a heterogeneous structure of the $l_o$ phase,
similar to a microemulsion, with ordered lipid nanodomains in
equilibrium with a disordered membrane was recently supported both
by theory and experiment. The computational work by Meinhardt, Vink
and Schmid~\cite{Meinhardt:2013} and Sodt {\em et al.}~\cite{Sodt:2014}
and the experimental papers by Armstrong {\em et
al.}~\cite{ArmstrongEBJ:2012,Armstrong:2013,Armstrong:2014} using
neutron diffraction were conducted using binary DPPC/cholesterol and
DMPC/cholesterol systems.

We combined coarse-grained molecular dynamics simulations including
20,000 lipid/cholesterol molecules with neutron diffraction using
deuterium labelled cholesterol molecules to study the cholesterol
structure in the liquid-ordered phase of DPPC bilayers. The
simulations present evidence for a heterogenous membrane structure
at 17~mol\% and 60~mol\% cholesterol and the formation of small,
transient domains enriched in cholesterol. The molecular structure
of the cholesterol molecules within these domains was determined by
neutron diffraction at 32.5~mol\% cholesterol. Three structures were
observed: (1) A fluid-like structure with strongly bound pairs of
cholesterol molecules as manifestation of the liquid-disordered
($l_d$) phase; (2) A highly ordered lipid/cholesterol phase where
the lipid/cholesterol complexes condense in a monoclinic structure,
in accordance with the umbrella model; and (3) triclinic cholesterol
plaques, i.e., cholesterol bilayers coexisting with the lamellar lipid
membranes.

\begin{figure}
\centering
\includegraphics[width=.87\columnwidth,angle=0]{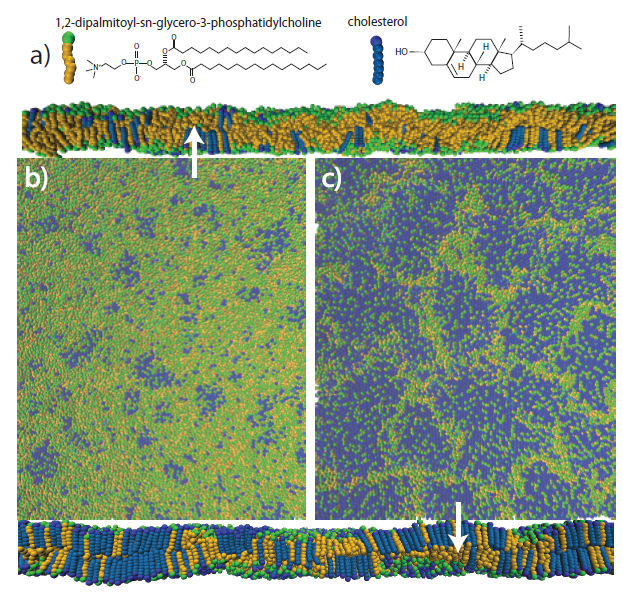}
\caption[]{a) Schematic representation of DPPC and cholesterol
molecules used in the simulations. b) Snapshot of the simulation at
$\mu$=8.5~$k_BT$, resulting in a cholesterol concentration of
$\approx$17~mol\%.  c) Snapshot of the simulation at
$\mu$=7.8~$k_BT$ resulting in a cholesterol concentration of
$\approx$60~mol\%.} \label{fig:simulationsnapshots}
\end{figure}

The simulations use a simple coarse-grained lipid model
\cite{SchmidDLW:2007} which reproduces the main phases of DPPC
bilayers including the nanostructured ripple phase $P_{\beta'}$
\cite{SchmidDLW:2007} and has similar elastic properties in the
fluid phase \cite{West:2009}. In this model, lipids are represented
by short linear chains of beads, with a `head bead' and several
`tail beads' (Fig.~\ref{fig:simulationsnapshots}~a)), which are
surrounded by a structureless solvent. The model was recently
extended to binary lipid/cholesterol mixtures. The cholesterol molecules are modelled shorter and stiffer
than DPPC, and they have an affinity to phospholipid molecules,
reflecting the observation that sterols in bilayers tend to
be solubilized by lipids \cite{Lindblom:2009}.
In our previous work,
we have reported on the behavior of mixed bilayers with small
cholesterol content~\cite{Meinhardt:2013}. Locally, phase{-}separation
was observed between a $l_o$ and a $l_d$ phase. On large scales,
however, the system assumes a two-dimensional microemulsion-type
state, where nanometer-sized cholesterol-rich domains are embedded
in a $l_d$ environment. These domains are stabilized by a
coupling between monolayer curvature and local ordering~\cite{Meinhardt:2013}, suggesting that raft
formation is closely related to the formation of ripples in
one-component membranes. In the following, we will discuss the
behavior of our model membranes at larger cholesterol concentrations
and discuss the implications for experiments.

The simulations were done at constant pressure, constant
temperature, and constant zero surface tension in a semi-grandcanonical
ensemble where lipids and cholesterol molecules can switch their
identity. The cholesterol content is thus driven by a chemical
potential parameter $\mu$. Simulation results are given in units of
$\sigma\approx 6$~\AA\ \cite{West:2009} and the thermal energy $k_B
T$. Typical equilibrated simulation snapshots (side view and top
view) are shown in Figs.~\ref{fig:simulationsnapshots}~b) and c). At
low cholesterol concentration ($\mu = 8.5~k_BT$), one observes small
rafts as discussed earlier. At higher cholesterol concentration
(lower $\mu$), the cholesterol-rich rafts grow and gradually fill up
the system, but they still remain separated by narrow
cholesterol-poor `trenches'. The side view shows that these
trenches have the structure of line defects where opposing
monolayers are connected. Such line defects are also structural
elements of the ripple phase in one-component bilayers
\cite{Vries:2005,Lenz:2007}.

\begin{figure}
\centering
\includegraphics[width=1.0\columnwidth,angle=0]{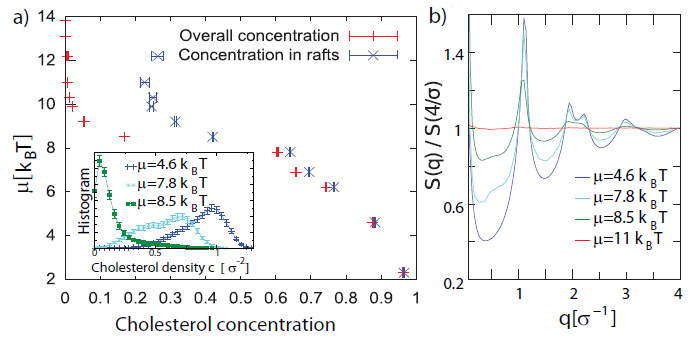}
\caption[]{a) Total cholesterol concentration and cholesterol
concentration inside rafts for different chemical potential $\mu$.
Inset shows a histogram of local cholesterol densities, taken using
squares of area $25 \sigma^2 \approx 9~\mbox{nm}^2$. b)
Radially averaged two-dimensional lateral structure factor of
cholesterol head groups for different $\mu$ as indicated. The level
of molecular order increases with decreasing $\mu$, i.e., increasing
cholesterol concentration.} \label{fig:simulationresults}
\end{figure}
With increasing cholesterol concentration, the structure of the
rafts changes qualitatively. This is demonstrated in
Fig.~\ref{fig:simulationresults}~a), which shows that the
cholesterol concentration inside rafts remains constant (around
25\%) for a range of chemical potentials $\mu > 8.5~k_B T$, but
then increases rapidly  at $\mu \le 8~k_B T$. Along with this
concentration increase, the peaks in the lateral structure factor of cholesterol
head groups in Fig.~\ref{fig:simulationresults}~b) become more pronounced, indicating a substantial increase in molecular
order. We should note that the coarse-grained model used in the
simulations is not suitable for studying details of the molecular
arrangement inside the ordered structures.
However, one can analyze the transition
between states with high and low $\mu$ by analyzing the distribution
of local cholesterol densities (Fig.~\ref{fig:simulationresults}~a),
inset). At high $\mu$, the histogram has a maximum at cholesterol
density $c$ close to zero and decays for higher $c$ with a
broad tail that reflects the contribution of the rafts. At low
$\mu$, it exhibits a marked maximum at $c \approx 1 \sigma^{-2}$,
corresponding to bilayer regions consisting purely of cholesterol.
In the intermediate regime, corresponding to the situation shown in
Fig.~\ref{fig:simulationsnapshots}~c), the histogram of cholesterol
densities features two broad peaks around $c {\approx} 0.4 \sigma^{-2}$ and
$c{\approx} 0.7 \sigma^{-2}$. In this regime, almost pure cholesterol
plaques coexist with regions having cholesterol
compositions that are close to those of rafts in
cholesterol-poor membranes (high $\mu$ limit in
Fig.~\ref{fig:simulationresults}~a)).

The experimental observation of the $l_o$ phase in a cholesterol
lipid binary mixture was initially reported by Vist and Davis
\cite{Vist:1990}. The quantitative determination of binary
lipid-cholesterol phase diagrams has remained elusive. In
phospholipid membranes, most studies report the $l_o$ phase at
cholesterol concentrations of more than 30~mol\% \cite{Marsh:2010}.
The formation of cholesterol plaques, phase{-}separated cholesterol bilayers coexisting with the membrane, was reported to occur at $\approx$37.5~mol\% cholesterol in model lipid membranes \cite{Barrett:2013}. That leaves a relatively small range of cholesterol concentrations in the experiment (between about 30-37.5~mol\%), where the $l_o$ phase can be studied. Phase{-}separation may be driven in experiments by certain boundary conditions, not present in computer simulations. The simulations in Fig.~\ref{fig:simulationresults} can, therefore, access a much larger range of cholesterol concentrations and by studying concentrations slightly lower and higher than the experimentally accessible range, the corresponding structures could be emphasized in the computer model.

\begin{SCfigure*}
\centering
\includegraphics[width=0.7\textwidth,angle=0]{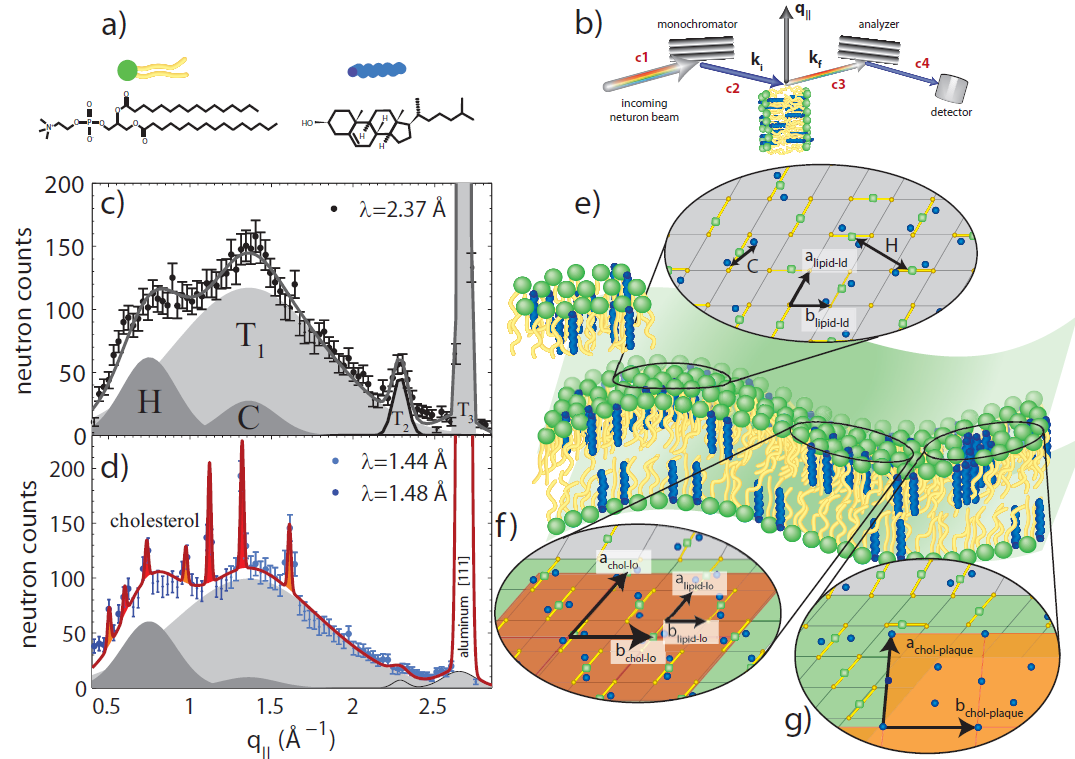}
\caption{a) Schematics of DPPC and (deuterated) cholesterol
molecules. b) Sketch of the scattering geometry. $q_{||}$ denotes the
in-plane component of the scattering vector. c) Diffraction measured
at $\lambda$=2.37~\AA\ showing broad, fluid-like peaks. d) Data
measured at $\lambda$=1.44 and 1.48~\AA. Several pronounced Bragg
peaks are observed in addition to the broad peaks in a). e) Cartoon
of the different molecular structures: Pairs of cholesterol
molecules in the liquid-disordered regions of the membrane in
equilibrium with highly ordered cholesterol structures such as the
umbrella structure f) and cholesterol plaques g).  An aluminum Bragg peak due to the windows of the humidity chamber and the sample holder is present at $q_{||}$ = 2.68~\AA $^{-1}$. Aluminum forms a face-centered cubic lattice with lattice parameter a= 4.04941~\AA \cite{Witt:1967}.
\label{Fig:NeutronInPlane}}
\end{SCfigure*}

We used neutron diffraction to measure the lateral cholesterol
structure in DPPC bilayers containing 32.5~mol\% at
$T$=50$^{\circ}$C and a D$_2$O relative humidity of $\approx$100\%,
ensuring full hydration of the membranes. Deuterium labeled
cholesterol (d7) was used such that the experiment was
sensitive to the arrangements of the cholesterol molecules.
Schematics of the two molecules are shown in
Fig.~\ref{Fig:NeutronInPlane}~a). Highly oriented, solid supported
membrane stacks on silicon wafers were prepared, as detailed in the
Supplementary Material. The sample was aligned in the neutron beam
such that the scattering vector, $\vec{Q}$, was placed in the plane
of the membranes (Fig.~\ref{Fig:NeutronInPlane}~b)). This in-plane
component of the scattering vector is referred to as $q_{||}$.

Two setups were used: a conventional high energy and momentum
resolution setup using a neutron wavelength of $\lambda$=2.37~\AA\
and a low energy and momentum resolution setup with smaller
wavelengths of $\lambda$=1.44 and 1.48~\AA. The latter setup was
reported to efficiently integrate over small structures and provide
a high spatial resolution capable of detecting small structures and
weak signals
\cite{Armstrong:2012a,Armstrong:2013,Rheinstadter:2013}. The two
setups could be readily switched during the experiment by changing
the incoming neutron wavelength, $\lambda$, without altering the
state of the membrane sample. Data taken using the conventional
setup are shown in Fig.~\ref{Fig:NeutronInPlane}~c) and display a
diffraction pattern with broad peaks, typical of a fluid-like
structure.

Peaks $T_1$, $T_2$ and $T_3$ in Fig.~\ref{Fig:NeutronInPlane}~c)
correspond to the hexagonal arrangement of the lipid tails with a
unit cell of $a_{\text{lipid}-l_d}=b_{\text{lipid}-l_d}$=5.58~\AA\
and $\gamma$=120$^{\circ}$, in agreement with Armstrong {\em et al.} \cite{Armstrong:2013}. By calculating the (coherent) scattering
contributions (Table~S3), cholesterol and lipid molecules contribute
almost equally to the scattering in the $l_d$ phase such that the
corresponding signals are observed simultaneously in
Fig.~\ref{Fig:NeutronInPlane}~c). Peak $H$ agrees well with an
average nearest neighbor head group-head group distance of
$\approx$8.4~\AA. Peak $C$ only occurs in the
presence of deuterated cholesterol molecules. It was, therefore,
assigned to a nearest neighbor distance of $\approx$4.6~\AA\ \sout{($\pm$0.2~\AA)}{($\pm$0.5~\AA)} of cholesterol molecules in
the $l_d$ phase, i.e., to pairs of strongly bound cholesterol
molecules, as shown in Fig.~\ref{Fig:NeutronInPlane}~e). Details of the fitting procedure are given in the Supplementary Material.
\begingroup
\squeezetable
\begin{table}
\centering
\begin{ruledtabular}
\begin{tabular}{c|c|c|c|c|c|c}
   \multirow{3}{*}{} & \multirow{2}{*}{Amplitude} & \multirow{2}{*}{Center} & \multirow{2}{*}{$\sigma_G$}  & \multirow{3}{*}{$l_d$} & monoclinic & triclinic \\
   & &  & & & cholesterol & cholesterol\\
   & (counts) & (\AA$^{-1}$)  & (\AA$^{-1}$) & & $l_o$-type structure & plaque\\ \hline
    \multirow{4}{*}{Fig.~\ref{Fig:NeutronInPlane}~c)} & 62&0.75&0.17& $H$ &&\\
    &117&1.360&0.46& $T_1$ &&\\
    &27.5&1.360&0.17& $C$ &&\\
    &46.6&2.289&0.05& $T_2$ &&\\
    &15.0&2.650&0.10& $T_3$ &&\\\hline
    \multirow{7}{*}{Fig.~\ref{Fig:NeutronInPlane}~d)}&19.8&0.5&0.01&&&[1 0 0]\\
    &34.8&0.55&0.01&&[1 $\bar{1}$ 0]\\
    &34.3&0.74&0.01&&[1 0 0]&[1 1 0]\\
    &33.5&0.98&0.01&&&[2 0 0]\\
    &110&1.12&0.01&&[2 $\bar{1}$ 0]&\\
    &117.8&1.32&0.01&&[1 1 0]&\\
    &60.0&1.61&0.01&&&[1 3 0]\\
\end{tabular}
\end{ruledtabular}
\caption{Peak parameters of the correlation peaks observed in
Fig.~\ref{Fig:NeutronInPlane}~c) and d) and the association with the
different cholesterol structures, such as $l_d$, $l_o$-type
structure and cholesterol plaque. $H$ and $C$ label the nearest neighbor distances of lipid head groups and cholesterol molecules, respectively; $T_1$, $T_2$, and $T_3$ denote the unit cell of the lipid tails in the $l_d$ regions of the membrane. Peaks
were fitted using Gaussian peak profiles and widths are listed as
Gaussian widths, $\sigma_G$.} \label{Table:PeakValues}
\end{table}
\endgroup

Several pronounced Bragg peaks are observed at neutron wavelengths
of $\lambda$=1.44 and 1.48~\AA\ in Fig.~\ref{Fig:NeutronInPlane}~d)
in addition to the broad correlation peaks. Due to the high
cholesterol concentration in $l_o$-type structures and plaques and
the scattering lengths of DPPC and $d$-cholesterol molecules, the
corresponding coherent scattering signal in
Fig.~\ref{Fig:NeutronInPlane}~d) is dominated by the deuterated
cholesterol molecules. As listed in Table~\ref{Table:PeakValues},
the peak pattern is well described by a superposition of two
2-dimensional structures: a monoclinic unit cell with lattice
parameters $a_{\text{chol-lo}}=b_{\text{chol-lo}}$=11~\AA\ and
$\gamma$=131$^{\circ}$ and a triclinic unit cell with
$a_{\text{chol-plaque}}=b_{\text{chol-plaque}}$=12.8~\AA\ and
$\gamma$=95$^{\circ}$ (the values for $\alpha$ and $\beta$ could not
be determined from the measurements but were taken from
\cite{Rapaport:2001,Barrett:2013} to be $\alpha=91.9^{\circ}$ and
$\beta=98.1^{\circ}$).

The lipid structure in the $l_o$-type structures in binary
DPPC/32.5~mol\% cholesterol bilayers was recently reported by
Armstrong {\em et al.} from neutron diffraction using deuterium
labelled lipid molecules \cite{Armstrong:2013}. The lipid tails were
found in an ordered, gel-like phase organized in a monoclinic unit
cell with $a_{\text{lipid-lo}}=b_{\text{lipid-lo}}$=5.2~\AA\ and
$\gamma$=130.7$^{\circ}$, as shown in Fig.~\ref{Fig:NeutronInPlane}~f).  The
cholesterol unit cell determined from the diffraction data in Fig.~\ref{Fig:NeutronInPlane}~c) is
indicative of a doubling of the lipid tail unit cell for the
cholesterol molecules. The corresponding cholesterol structure consists of
cholesterol pairs alternating between two different orientations.

The $l_d$ and the $l_o$-type structures can be related to the
well-known umbrella model \cite{Huang:1999}, where one lipid
molecule is assumed to be capable to `host' two cholesterol
molecules, which leads to a maximum cholesterol solubility of
66~mol\% in saturated lipid bilayers. In this scenario the term
umbrella model refers to two cholesterol molecules closely
interacting with one lipid molecule. Cholesterol plaques, i.e.,
cholesterol bilayers coexisting with the lamellar membrane phase
were reported recently by Barrett {\em et al.}~\cite{Barrett:2013}
in model membranes containing high amounts of cholesterol, above
40~mol\% for DMPC and 37.5~mol\% for DPPC. The triclinic peaks in
Fig.~\ref{Fig:NeutronInPlane}~d) agree well with the structures
published and were, therefore, assigned to cholesterol plaques.

Hence both coarse-grained molecular simulations and
neutron diffraction data suggest the coexistence of a liquid
disordered membrane with two types of highly ordered cholesterol
structures: One with some lipid content (Fig.~\ref{Fig:NeutronInPlane}~f), corresponding to
the first shoulder in the density histogram at $\mu=7.8 k_B T$ (Fig.~\ref{fig:simulationresults}~a),
inset), and one almost exclusively made of cholesterol (Fig.~\ref{Fig:NeutronInPlane}~g),
corresponding to the second peak at $\mu = 7.8 k_B T$ in Fig.~\ref{fig:simulationresults}~a).
The existence of these structures in the experiment should be robust in binary systems and not depend on, for instance, the sample preparation protocol \cite{Elizondo:2012}.

The neutron diffraction data present evidence for pairs of strongly
bound cholesterol molecules. We
note that the scattering experiment was not sensitive to {\em
single} cholesterol molecules, however, the formation of cholesterol
dimers with a well defined nearest neighbor distance leads to a
corresponding peak in the data in Fig.~\ref{Fig:NeutronInPlane}~c)
and d). An attractive force between cholesterol molecules in a POPC
bilayer and the formation of cholesterol dimers was reported from MD
simulations \cite{Andoh:2012}. Such a force is likely related to the
formation of lipid/cholesterol complexes \cite{McConnell:2003} and
the umbrella model. However,
it is not straightforward to estimate the percentage of dimers from
the experiments. A dynamical equilibrium between dimers and monomers
is a likely scenario~\cite{Dai:2010}.

The dynamic domains observed in this study are not biological rafts,
which are thought to be more complex, multi-component structures in
biological membranes. In the past, domains have been observed in
simple model systems, but only those designed to be `raft-forming'
mixtures.  In these cases the domains that form are stable
equilibrium structures, and are not likely related to the rafts that
exist in real cells~\cite{Rheinstadter:2013}. The small and
fluctuating domains observed in binary systems may be more closely
related to what rafts are thought to be~\cite{Simons:2010}, and are potentially the
nuclei that lead to the formation of rafts in biological membranes.
The characteristic overall length scale for nanodomains in the
simulations is around 20$\sigma$, corresponding to 10-20~nanometers.
Both simulations and experiments indicate that there are two types
of cholesterol-rich patches coexisting with cholesterol-poor
liquid-disordered regions, i.e., ordered $l_o$-type regions
containing lipids and cholesterol and cholesterol plaques. The
transition between these two is gradual in the coarse-grained
simulations. In real membranes, they have different local structure
(monoclinic in $l_o$-type regions, triclinic in plaque regions),
which may stabilize distinct domains.

\acknowledgements This work was supported by the German Science
Foundation within the collaborative research center SFB-625.
Simulations were carried out at the John von Neumann Institute for
Computing (NIC) J\"ulich and the Mogon Cluster at Mainz University.
Experiments were funded by the Natural Sciences and Engineering
Research Council (NSERC) of Canada, the National Research Council
(NRC), the Canada Foundation for Innovation (CFI), and the Ontario
Ministry of Economic Development and Innovation. L.T.~is the
recipient of a Canada Graduate Scholarship, M.C.R.~is the recipient
of an Early Researcher Award from the Province of Ontario.
\bibliography{Membranes}


\clearpage
\newpage
\onecolumngrid
\renewcommand{\baselinestretch}{1.50}\normalsize
\setcounter{section}{0} \setcounter{subsection}{0}
\setcounter{subsubsection}{0} \setcounter{figure}{0}
\setcounter{page}{1}
\renewcommand{\thepage}{S\arabic{page}}
\renewcommand{\thesection}{S\arabic{section}}
\renewcommand{\thesubsection}{S\arabic{section}.\arabic{subsection}}
\renewcommand{\thesubsubsection}{S\arabic{section}.\arabic{subsubsection}}
\renewcommand{\thetable}{S\arabic{table}}
\renewcommand{\thefigure}{S\arabic{figure}}

\section*{\Large Supplementary Material to:\protect\\The Structure of Cholesterol in Lipid Rafts}
\section{Materials and Methods}

\subsubsection{Coarse-Grained Simulation Model\label{Sec:Simulations}}
The model is defined in terms of the length unit
$\sigma\approx$0.6~nm and the energy unit $\epsilon\approx 0.36\cdot
10^{-20}$~J \cite{West:2009}. Phospholipid molecules ($P$) are
represented by simple flexible chains of beads with a hydrophilic
head and a hydrophobic tail, which self-assemble in the presence of
structureless solvent beads \cite{Lenz:2005}. Cholesterol molecules
($C$) have the same basic structure, but they are shorter and
stiffer except for one flexible end. All lipids are linear chains of
six tail beads attached to one head bead, connected by finite
extension nonlinear elastic (FENE) springs with the spring constant
$k_b =100\frac{\epsilon}{\sigma^2}$, equilibrium bond lengths $r_0 =
0.7\sigma$ ($P$ lipid) and $r = 0.6\sigma$ ($C$ lipid), and
logarithmic cutoffs at $\Delta r_{\text{max}} = 0.2\sigma$ ($P$) and
$\Delta r_{\text{max}} = 0.15\sigma$ ($C$).
We have established in previous work that membranes containing
pure P lipids roughly reproduce the behavior of DPPC bilayers
\cite{Lenz:2007, West:2009}. Fully stretched C-lipid chains have
a length of $3.6 \sigma$, corresponding to $\sim$ 22 \AA,
which is reasonably close to the estimated length
$\sim$ 19~\AA\ of cholesterol molecules \cite{Nes:2011, HostaRigau:2013}.
Consecutive bonds in a
chain with angle $\theta$ are subject to a stiffness potential
UBA($\theta$) = $k_{\theta}(1-\cos(\theta))$ with stiffness constant
$k_{\theta} = 4.7\epsilon$ ($P$ lipids), $k_{\theta} =
100\epsilon$ ($C$ lipids, first four angles), and $k_{\theta} =
4.7\epsilon$($C$ lipid, last angle). Beads that are not directly
bonded with each other interact via a Lennard-Jones potential
$U_{LJ}(r/\zeta)=\epsilon_{LJ}\left(\left(\frac{\zeta}{r}\right)^{12}
-2\left(\frac{\zeta}{r}t\right)^6\right)$, which is truncated at a
cutoff radius $r_c$ and shifted such that it remains continuous. At
$r_c$ = 1, one recovers the purely repulsive Weeks-Chandler-Anderson potential \cite{Schmid:2007}. The interaction parameters for
pairs of $P$ or $C$ beads (head or tail) and solvent beads are
taken from \cite{Meinhardt:2013} and given in
Table~\ref{Table:theory}.

\begin{table}[H]
\begin{ruledtabular}
\begin{tabular}{l|c|c|c}
   Bead type-bead type & $\epsilon/\epsilon$ & $\zeta/\sigma$ & $r_c/\zeta$\\\hline
    Head(any)-head(any) & 1.0 & 1.1 & 1.0\\
    Head(any)-tail(any) & 1.0 & 1.05 & 1.0\\
    Head(any)-solvent & 1.0 & 1.1 & 1.0\\
    Tail($P$)-tail($P$) & 1.0 & 1.0 & 2.0\\
    Tail($P$)-tail($C$) & 1.0 & 1.0 & 2.0\\
    Tail($C$)-tail($C$) & 0.9 & 1.0 & 2.0\\
    Tail(any)-solvent & 1.0 & 1.05 & 1.0\\
    Solvent-solvent&0 & 0 & 0\\
\end{tabular}
\end{ruledtabular}
\caption{Interaction parameters of the coarse-grained molecular
dynamics simulations.\label{Table:theory}}\end{table}

Hence, all non-bonded interactions except the tail-tail interactions
are repulsive, and the attraction between $C$ tail beads is weaker
than that between other tail beads, which effectively leads to
an enhanced affinity between $C$-chains and $P$-chains.
The model was studied by Monte
Carlo simulations at constant pressure $P = 2\epsilon/\sigma^3$ and
constant zero surface tension in a fluctuating box of variable size
and shape \cite{Schmid:2007}. The total number of lipids was kept
fixed, but the composition was allowed to fluctuate (semi-grand
canonical ensemble). This was implemented by means of
configurational bias Monte Carlo moves \cite{Frenkel:2001}, during
which the identity of a lipid was switched between $P$ and $C$.
The use of the semi-grand canonical ensemble was necessary to
ensure that the configurations could be equilibrated and that the
finite domains were not simply the result of an incomplete phase
separation. Even in the semi-grand canonical ensemble, the equilibration
times were still of order 1-3~million Monte Carlo sweeps.
Simulations were carried out at the John von Neumann Institute for
Computing (NIC) J\"ulich and the Mogon Cluster at Mainz University.

\subsubsection{Sample Preparation and Neutron Diffraction Experiment}
Deuterated cholesterol (d7) was used to enhance the intensity of the
 out-of-plane and in-plane neutron Bragg diffraction peaks.
Highly oriented, multi-lamellar stacks of
1,2-dipalmitoyl-sn-glycero-3-phosphocholine (DPPC) with 32.5~mol\%
cholesterol-d7 were prepared on 2$^{\prime\prime}$ single-side
polished Si wafers of thickness 300~$\mu$m. A solution of 16.7~mg/mL
DPPC with 32.5~mol\% cholesterol in 1:1 chloroform and
2,2,2-trifluoroethanol (TFE) was prepared. The Si wafers were
cleaned by alternate 12 minute sonications in ultra pure water and
methanol at 313~K. This process was repeated twice.  The cleaned
wafers were placed on a heated sample preparation surface, which was
kept at 50$^{\circ}$C. This temperature is well above the main phase
transition for DPPC, thus the heated substrates ensured that the
lipids were in the fluid phase after deposition. 1.2~mL of the lipid
solution was deposited on each Si wafer and allowed to dry. The
wafers were kept under vacuum overnight to remove all traces of the
solvent. Samples were then hydrated with heavy water, D$_2$O, and
annealed in an incubator at 328~K for 24~hours. Following this
protocol, each wafer contained $\approx$3,000 highly oriented
stacked membranes with a total thickness of $\approx$10~$\mu$m.
Sixteen such Si wafers were stacked with two 0.3~mm aluminum spacers
placed in between each wafer to allow for the membranes to be
properly hydrated. The ``sandwich'' sample was kept in a sealed
temperature and humidity controlled aluminum chamber. Hydration of
lipid membranes from water vapor was achieved by independently
adjusting the temperature of the heavy water reservoir, the sample
and the chamber cover. Temperature sensors were installed close to
the sample. A water bath was used to control the temperature of the
D$_2$O reservoir, and the temperatures of the sample
and its cover were controlled using Peltier elements. 

\begin{figure}[h]
\centering
\includegraphics[width=0.70\columnwidth,angle=0]{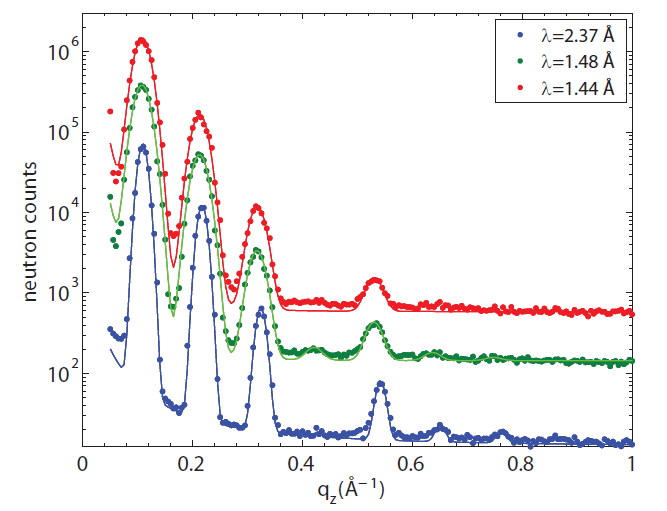}
\caption{Neutron diffraction data for each scattering wavelength used to
probe the out-of-plane structure along the membrane normal, $q_z$. The wavelength 2.37~\AA\ is the
wavelength used in conventional neutron scattering setups, while
wavelengths 1.48~\AA\ and 1.44~\AA\ were used in for the low energy,
low $Q$ resolution setups.\label{Fig:Refl}}
\end{figure}

\subsubsection{Sample Characterization}
The sample was mounted vertically in the neutron beam such that the
scattering vector, $Q$, could either be placed in the plane of the
membranes ($q_{||}$), or perpendicular to it ($q_{z}$), as shown in
Fig.~3~b). The out-of-plane and in-plane structures could be
measured by simply rotating the sample by 90$^{\circ}$. The lamellar
spacing, $d_z$, i.e., the distance between two neighboring membranes
in the stack, was determined from neutron diffraction along the membrane normal, $q_z$. The
corresponding scans are shown in
Fig.~\ref{Fig:Refl} and show a series of pronounced and equally
spaced Bragg peaks indicative of uniform and well
developed lamellar membrane stacks. A concentration of 32.5~mol\%
cholesterol was previously carefully checked by Armstrong {\em et
al.} and confirmed to be in the liquid-ordered state for DPPC membranes
\cite{Armstrong:2013}. Scans were measured for different neutron
wavelengths, $\lambda$, and over the duration of the experiment of
about 3 weeks. The bilayer spacing stayed within the range of
57.8~\AA\ and 59.2~\AA, corresponding to a hydration of the bilayers
to better than 99.6\% \cite{Chu:2005}. The main transition
temperature, $T_m$, of DPPC is reported to be
$T$~=~314.4~K~\cite{Mabrey:1976,KatsarasPRE:1997}. All measurements
reported here were done at $T$~=~323.2~K (50$^{\circ}$C), well above
$T_m$.

\subsubsection{Scattering Cross Sections of the Different Molecular Components}
To determine the contribution of lipid and cholesterol molecules to
the scattering signals, the corresponding scattering lengths were
calculated and are listed in Table~\ref{Table:scatteringlength}.
Scattering lengths for protonated and deuterated lipid, cholesterol
and water molecules are included for completeness. Scattering from
lipid molecules dominates the scattering signal in a
DPPC-d62/32.5~mol\% cholesterol system. Lipid and cholesterol
molecules have equal scattering contributions in the
liquid-disordered phase at DPPC/66~mol\% cholesterol-d7 and both
signals were observed in the data in Fig.~3~c). However, cholesterol
molecules dominate at higher concentrations of, e.g., DPPC/66~mol\%
cholesterol-d7 in $l_o$-type or plaque structures (as seen in
Fig.~3~d)).
\begin{table}[H]
\begin{ruledtabular}
\begin{tabular}{c|c|c|c}
        &Molecule & Chemical Formula&Scattering Length (fm)\\\hline
        &DPPC & C$_{40}$H$_{80}$NO$_8$P & 27.63 \\
        &DPPC-d62 & C$_{40}$H$_{18}$D$_{62}$NO$_8$P & 673.1 \\
        &Cholesterol &C$_{27}$H$_{46}$O  & 13.25\\
        &Cholesterol-d7 &C$_{27}$H$_{39}$D$_7$O  & 86.12\\
        &Water & H$_2$O & -1.68\\
        &Heavy Water & D$_2$O & 19.15\\\hline
        \multirow{2}{*}{DPPC/32.5~mol\% Cholesterol-d7} &DPPC & C$_{40}$H$_{80}$NO$_8$P & 18.65 \\
            & Cholesterol-d7 &C$_{27}$H$_{39}$D$_7$O & 27.99\\\hline
        \multirow{2}{*}{DPPC/66~mol\% Cholesterol-d7} &DPPC & C$_{40}$H$_{80}$NO$_8$P & 9.11 \\
            & Cholesterol-d7 &C$_{27}$H$_{39}$D$_7$O & 56.84\\
\end{tabular}
\end{ruledtabular}
\caption{Scattering lengths of the different membrane
components.\label{Table:scatteringlength}}\end{table}

\subsubsection{Neutron Diffraction Experiment}
Experiments were conducted using the N5 triple-axis spectrometer at
the Canadian Neutron Beam Centre (Chalk River, ON, Canada).  The
three axes of the spectrometer refer to the axis of rotation of the
monochromator, the sample and the analyzer. The incident and final
neutron energies are defined by the Bragg reflections from pyrolytic
graphite (PG) crystals.  The divergence of the neutron beam was
controlled by Soller collimators. A schematic of the instrument's
configuration is shown in Fig.~3~b). The instrumental parameters for
the two setups used in this experiment are listed in
Table~\ref{Table:expsetups}. Energy and $Q$-resolution (given as
FWHM) were calculated using the ResLib software package by
A.~Zheludev \cite{reslib} adapted to the N5 spectrometer.
\begin{table}[H]
\begin{ruledtabular}
\begin{tabular}{c|c|c|c}
        $\lambda$ (\AA)&    E (meV)  & $\Delta E$ (meV)& $\Delta Q$ (\AA$^{-1}$)\\\hline
        2.37 & 14.6 & 0.757 & 0.020\\
        1.44 & 39.5 & 3.521 & 0.033\\
        1.48 & 37.3 & 3.239 & 0.032
\end{tabular}
\end{ruledtabular}
\caption{Instrumental parameters of the triple-axis
spectrometer.\label{Table:expsetups}}
\end{table}

The $\Delta E$ and $\Delta Q$ of a neutron triple-axis spectrometer
are determined by: (1)~the incident energy of the neutron beam;
(2)~the divergence of the neutron beam; and (3)~the wavelength
resolution of the monochromator and analyzer. Collimation was kept
constant during the course of the experiment and set to
(c1-c2-c3-c4): 30-18-28-60 (in minutes). Small and large $\Delta E$
setups were achieved by varying the incident energy of the incoming
neutrons. The longitudinal coherence length of the neutron beam,
$\xi$, is defined by $\xi=\lambda^2/\Delta\lambda$
\cite{Rauch:1993}. For a neutron spectrometer with incident neutron
energy $E$ and instrumental energy resolution $\Delta E$, $\xi$ can
be estimated to be $\xi\approx 18\sqrt{E}/\Delta E$
\cite{Armstrong:2012a}, where $E$ and $\Delta E$ are in meV. 

Note that the reason for the typically low monochromaticity of
neutron beams is to avoid further compromising the already low-flux
``white'' neutron beam, a situation that is very different for
synchrotron X-rays. Switching between the high and low energy
resolution setups was done by changing instrumental settings of the
neutron triple-axis spectrometer, which has an effect on $\Delta Q$
and $\Delta E$ of the beam. A smaller neutron wavelength leads to
strongly relaxed $\Delta Q$ and $\Delta E$. In addition, the
longitudinal coherence length of the neutron beam decreases. The
most significant changes between the high and low energy resolution
setups are: (1) a more efficient integration over smaller $q_{||}$
ranges to enhance small signals; and (2) a reduction of the
coherently added scattering volume.

\subsubsection{Details of the Fitting of In-Plane Diffraction Data}
\begin{table}[H]
\begin{ruledtabular}
\begin{tabular}{c|c|c|c}
        $\lambda$ (\AA)&    Lorentzian Amplitude (counts)  & Lorentzian Width (\AA$^{-1}$) & Constant Offset (counts)\\\hline
        2.37 & 730 & 0.45 & 135\\
        1.44 & 1060 & 0.29 & 170\\
        1.48 & 1000 & 0.29 & 175
\end{tabular}
\end{ruledtabular}
\caption{Parameters of the background used to fit the in-plane diffraction data. The incoherent scattering contribution was accounted for by a Lorentzian peak shape centered at $q_{||}$=0~\AA$^{-1}$.\label{Table:background}}
\end{table}
The in-plane diffraction data in Figure~3~c) and d) were fit by using a series of Gaussian peak shapes to determine the scattering contributions of lipid and cholesterol molecules. The high total scattering of the sample, due to the large percentage of protonated material, resulted in a large constant, $Q$-independent background. While small collimation was used, beam size was 2 inches by 2 inches to optimally illuminate the silicon wafers, leading to a significant scattering contribution at small Q-values, close to the direct beam. The background was accounted for by a Lorentzian peak centered at $q_{||}$=0~\AA$^{-1}$ including a constant. Details of the background for the different wavelengths used are given in Table~\ref{Table:background}. In contrast to coherent scattering, incoherent scattering is isotropic, $Q$-independent and well accounted for by a constant background at larger $Q$-values of $Q\gtrsim$0.45~\AA$^{-1}$.
\begin{figure}
\centering
\includegraphics[width=1.00\columnwidth,angle=0]{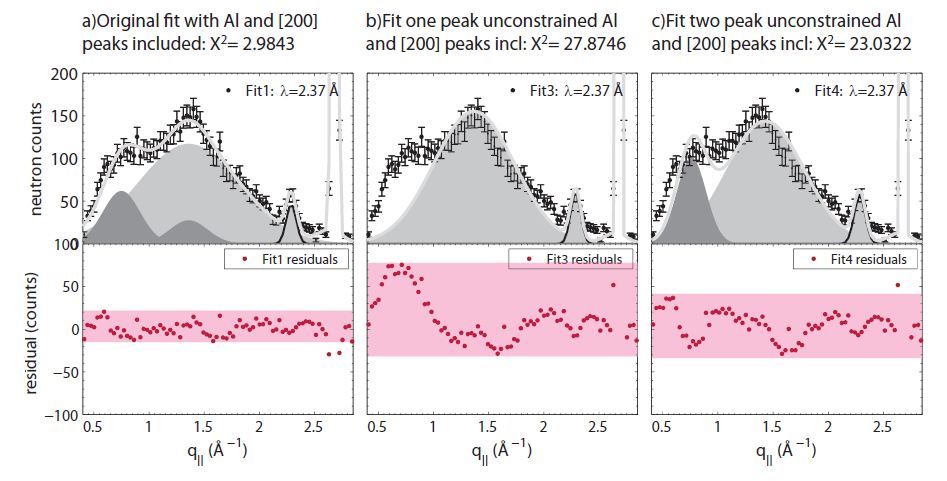}
\caption{Comparison of fitting different models to the in-plane diffraction data (Fig.~3 in the main paper). The fit including Gaussians peak profiles from lipid tails and head groups and cholesterol best describes the data with a significantly better $\chi^2$-value. \label{Fig:Comparison}}
\end{figure}

The in-plane diffraction data were tentatively fit with only lipid tail peaks, lipid tail and head group peaks, and lipid tail, head group and cholesterol peaks, as shown in Fig.~\ref{Fig:Comparison}. The corresponding residuals and  $\chi^2$-values are given in the figure. Based on the comparison, the model suggested in the main paper including contributions from lipid tails, head groups and cholesterol best describes the experimental data, with a significantly smaller $\chi^2$-value.

\begin{figure}
\centering
\includegraphics[width=1.00\columnwidth,angle=0]{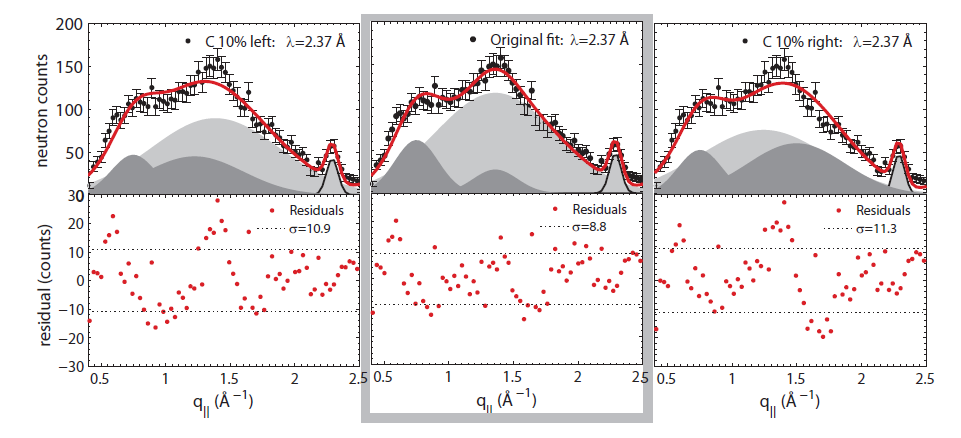}
\caption{{The uncertainty in determining the position of the cholesterol peak ($C$) can tentatively be estimated by shifting and constraining the peak away from its fitted position and evaluating the quality of the observed fit. The residuals ($(y^{data}_i-y^{fit}_i)-\bar{y}$) are plotted below each fit and the corresponding standard deviations, $\sigma$, are listed in the legend. $\sigma$ was found to increase by $\approx$25\% when the when the $C$ peak position was changed by more than \sout{5\%}{10\%}.}. \label{Fig:Cpeak}}
\end{figure}

{The uncertainly in the determination of the position of the cholesterol peak (the $C$-peak in Fig.~3~c) can tentatively be estimated by fixing the position of the cholesterol peak ($C$) and shifting it by \sout{5\%}$\pm$~10\% with respect to the \sout{fitted position}$C$ peak position in the original fit. \sout{Amplitudes of all peaks}All peak amplitudes were allowed to vary, however, all peak locations were fixed except for the lipid tail ($T_1$) peak. The result of this procedure is shown in Fig.~\ref{Fig:Cpeak}. In order to estimate the quality of the fitted curve, the mean (see Eq.~(\ref{Eq:mean}) from \cite{Taylor:1982}) and standard deviation of the residuals (Eq.~(\ref{Eq:stdd}) from \cite{Taylor:1982}) were calculated:}

\begin{equation}\label{Eq:mean}
 { \bar{y} =  \frac{\sum_{i=1}^n{(y^{data}_i-y^{fit}_i)}}{n} \;\;\;\mbox{and}}
\end{equation}
\begin{equation}\label{Eq:stdd}
 { \sigma = \sqrt{ \frac{\sum_{i=1}^n{\left((y^{data}_i-y^{fit}_i)-\bar{y}\right)^2}}{n-1} } ,}
\end{equation}

{where $\bar{y}$ is the mean of the deviation of the data from the fit, $\sigma$ is the standard deviation from the mean, and $n$ is the number of data points used in the calculation. The result is shown in Fig.~\ref{Fig:Cpeak}. The residuals and standard deviation from (Eq.~(\ref{Eq:stdd})) are plotted below the fit. Both the fits where the $C$ peak is fixed to $q_{||}$=1.22~\AA\ (equivalent to a shift of 10\% to the left) and where the $C$ peak is fixed to $q_{||}$=1.50~\AA\ (equivalent to a 10\% to the right) show a significant increase in the standard deviation. We note that the data range used in these calculations does not include the aluminum peak as its high intensity drowns out the quantitative differences in the aforementioned fits.}

{The analysis can be summarized as follows: the in-plane diffraction data are best described by peaks related to ordering of lipid tails, lipid head groups and cholesterol molecules. Based on the above analysis the uncertainty in the correlation distance between cholesterol molecules in a cholesterol-lipid-cholesterol complex \sout{on}in the liquid disordered phase was tentatively estimated to be \sout{5\%}{10\%}, resulting in 4.6~$\pm$~\sout{0.2}0.5~\AA.}


\end{document}